\documentclass{appolb}
\usepackage{graphicx}
\usepackage[T1]{fontenc}
\begin{document}
\title{Physics with next generation neutrino experiments: ESSnuSB%
\thanks{Presented at  ``Matter to the deepest: Recent developments in physics of fundamental interactions, XLVI international conference of theoretical physics (MTTD 2025)", University of Silesia in Katowice, Poland 15 - 19 September 2025}% 
}
\author{Monojit Ghosh \\
for the ESSnuSB Collaboration
\thanks{mghosh@irb.hr}
\address{Center of Excellence for Advanced Materials and Sensing Devices, Ru{\dj}er Bo\v{s}kovi\'c Institute, 10000 Zagreb, Croatia}
\\
}
\maketitle
\begin{abstract}

 In this proceedings we explore the physics potential of the ESSnuSBplus setup to study beam and non-beam based physics scenarios in both standard and new physics cases. The ESSnuSBplus setup consists of three neutrino sources: the main ESS linac, a low energy monitored neutrino beam and a low energy nuSTORM facility  and three detectors: the main far detector and two near detectors. The goal of this facility is to measure the leptonic CP phase with extremely high precision and the neutrino nucleus cross-section in the few hundred MeV region. 

\end{abstract}
  
\section{Introduction}

While the aim of the current generation long-baseline experiments \cite{T2K:2025wet} T2K and NO$\nu$A is to find the hint of the correct neutrino mass ordering, the true octant of the atmospheric mixing angle $\theta_{23}$ and the true value of the leptonic CP phase $\delta_{\rm CP}$, the goal of the future generation experiments T2HK \cite{Hyper-Kamiokande:2018ofw}, DUNE \cite{DUNE:2020ypp} and ESSnuSB \cite{Alekou:2022emd} is to confirm these hints in a firm footing. Apart from measuring the parameters in the standard  three flavour oscillation scenario, these experiments are also capable in probing various new physics scenarios. In this proceedings, we will study the capability of the ESSnuSBplus setup \cite{ESSnuSB:2024tmn} to probe a wide variety of physics cases in both standard and new physics cases.

The text is organized as follows. In the next section we will describe the ESSnuSBplus set up in detail and then in the following sections we demonstrate its sensitivity to various physics cases. Finally we will summarize the results and conclude.

\section{The ESSnuSBplus facility}
\label{3nu}

\begin{figure}[htb]
\centerline{
\includegraphics[width=15cm]{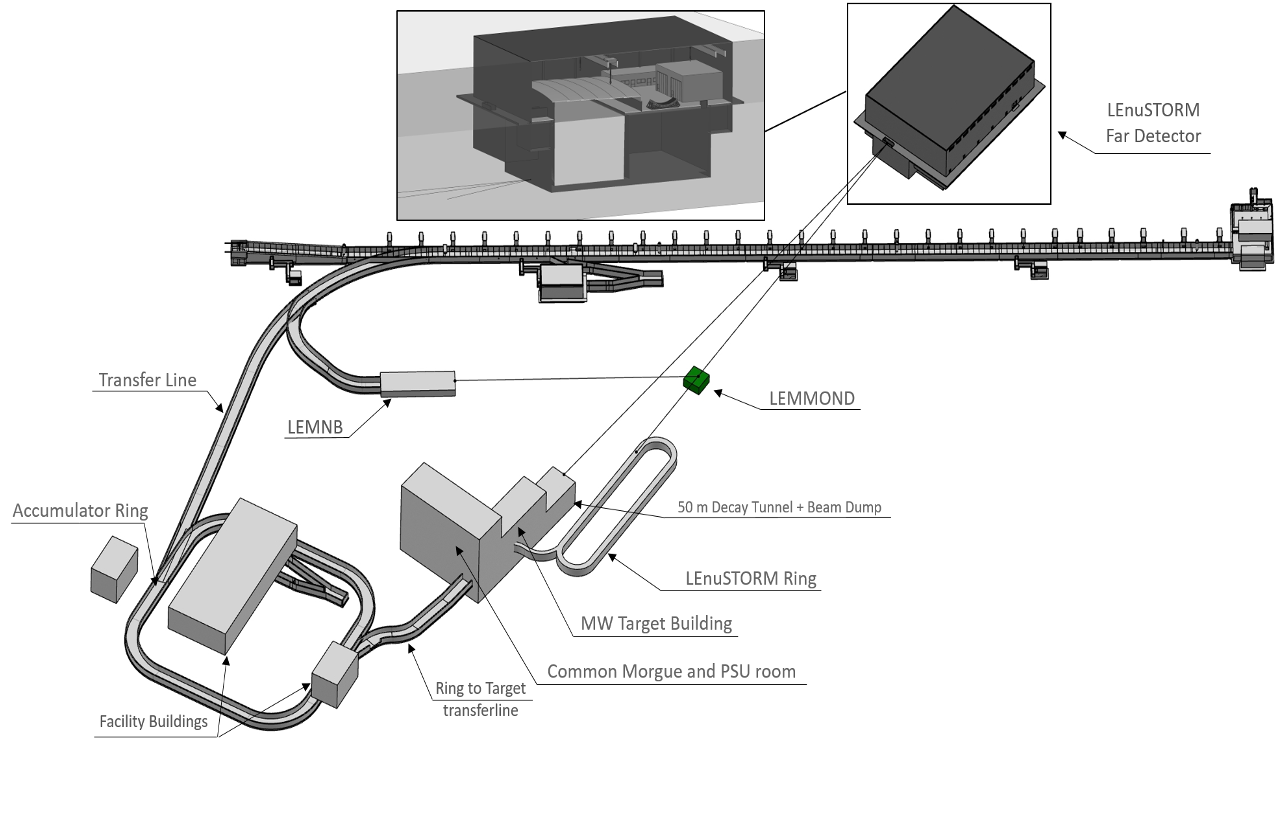}}
\caption{The ESSnuSBplus facility.}
\label{fig1}
\end{figure}

The ESSnuSBplus facility in Sweden will consist of three neutrino sources and three detectors. A preliminary layout of the facility has been shown in Fig.~\ref{fig1}. The main ESS linac in Lund will deliver a 5 MW proton beam with 2.5 GeV proton energy and 2.8 ms pulse. Then the accumulator ring will reduce the pulse length to 1.2 $\mu$s in order to minimize atmospheric background. These protons will collide with a target to produce a intense beam of muon neutrinos via pion decay. These neutrinos will pass through a near detector located at 250 m (END) and then they will finally reach the far detector (FD) consisting of 540 kt ultra-pure water, located at a distance of 360 km at Zingruvan. In addition, this facility will also have a low energy monitored neutrino beam (LEMNB) which is inspired by the original ENUBET \footnote{to be implemented in nuSCOPE \cite{Acerbi:2025wzo}.} idea \cite{ENUBET:2025qhg}. This will be an instrumented decay pipe to measure neutrino flux to 1\% uncertainty. These  neutrinos will be detected at another near detector called LEMMOND. Moreover, for the measurement of cross-section, this facility will also have a low energy neutrino source from muon decay (LEnuSTORM) inspired by the original nuSTORM idea \cite{nuSTORM:2025tph}. These neutrinos will be detected in LEMMOND and END detectors. 

\section{Results}

\subsection{Sensitivity to $\delta_{\rm CP}$}

\begin{figure}[htb]
\centerline{
\includegraphics[width=7.5cm]{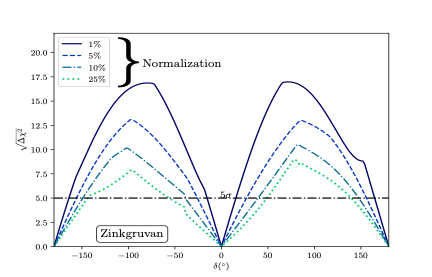}
\includegraphics[width=7.5cm]{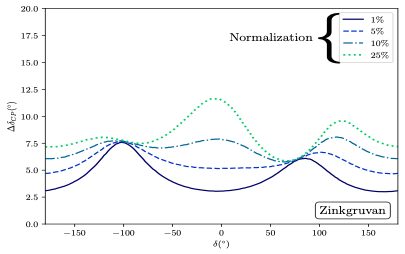}}
\caption{CP sensitivity of the ESSnuSB experiment.}
\label{fig2}
\end{figure}

Fig.\ref{fig2} shows the CP sensitivity of the ESSnuSB facility using the main ESS beam, the END and the FD. The left panel shows the CP violation sensitivity whereas the right panel shows the CP precision sensitivity. The different curves in each panel show the different values of systematic errors. From the plot we see that ESSnuSB will have world leading sensitivity for the $\delta_{\rm CP}$. The CP sensitivity of ESSnuSB is better than T2HK and DUNE because ESSnuSB probes the second oscillation maximum in the probability spectrum whereas the other experiments probe the first oscillation maxima. As the CP sensitivity at the second oscillation maximum is three times higher as compared to the first maximum, ESSnuSB provides a better sensitivity. 

\subsection{Sensitivity to sterile neutrinos}

\begin{figure}[htb]
\centerline{
\includegraphics[width=15cm]{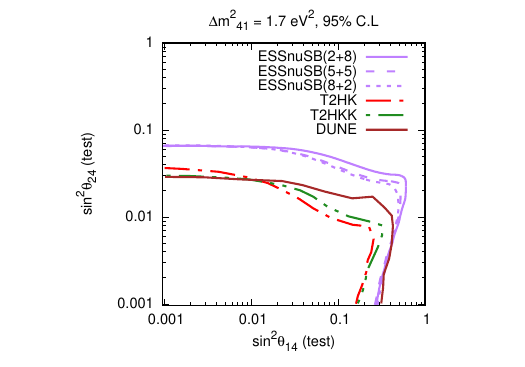}}
\caption{Sensitivity of the ESSnuSB experiment to sterile neutrinos.}
\label{fig3}
\end{figure}

Fig.\ref{fig3} shows the sensitivity of the ESSnuSB facility to the eV scale sterile neutrinos using the main ESS beam and the FD \cite{Ghosh:2019zvl}. In this figure we have compared the sensitivity of ESSnuSB experiment to constrain the sterile mixing parameters $\theta_{14}$ and $\theta_{24}$ with T2HK and DUNE. The different curves for ESSnuSB refer to different run-times in neutrino and antineutrino mode. These results were generated using an older version of the flux files and older version of the detector responses. We plan to update these results with our latest flux and detector responses. This plot shows results only for the far detector, whereas we expect significant change in sensitivity once we include the near detectors in the analysis. 

\subsection{Sensitivity to neutrino decay}

\begin{figure}[htb]
\centerline{
\includegraphics[width=10cm]{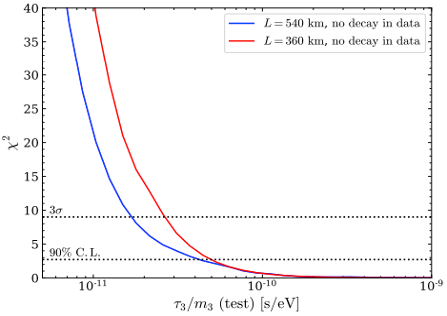}}
\caption{Sensitivity of the ESSnuSB experiment to invisible neutrino decay.}
\label{fig4}
\end{figure}

Fig.\ref{fig4} shows the sensitivity of the ESSnuSB facility to invisible neutrino decay using the main ESS beam and the FD \cite{Choubey:2020dhw}. This plot shows the capability of the ESSnuSB FD to constrain the lifetime of the heavier neutrino mass state $m_3$ in the normal mass ordering of the neutrinos. The red curve is for the baseline of 360 km whereas the blue curve is for the baseline of 540 km which was one of the earlier baseline options of this experiment. While comparing our sensitivity with the other experiments, we find our sensitivity to be better than DUNE. 

\subsection{Sensitivity to scalar NSI}

\begin{figure}[htb]
\centerline{
\includegraphics[width=10cm]{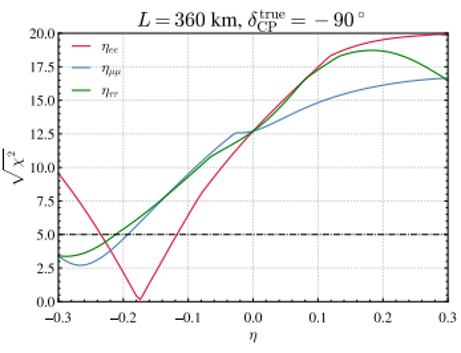}}
\caption{Sensitivity of the ESSnuSB experiment to scalar NSI.}
\label{fig5}
\end{figure}

Fig.\ref{fig5} shows the sensitivity of the ESSnuSB facility to non-standard interaction mediated by a scalar field (SNSI) using the main ESS beam and the FD \cite{ESSnuSB:2023lbg}. This figure shows the CP violation discovery $\chi^2$ as a function of SNSI parameter $\eta$ for $\delta_{\rm CP} = -90^\circ$. In this figure, the different curves show sensitivity for three diagonal SNSI parameters. The point $\eta = 0$ refers to the standard three flavour case where all the curves coincide. From this figure we can see an interesting phenomenon that for $\eta_{ee} = -1.8$, the CP violation sensitivity of the ESSnuSB experiment vanishes. Therefore if SNSI exists in nature, then this can be a clear signal of this theory. 

\subsection{Sensitivity to quantum decoherence}

\begin{figure}[htb]
\centerline{
\includegraphics[width=10cm]{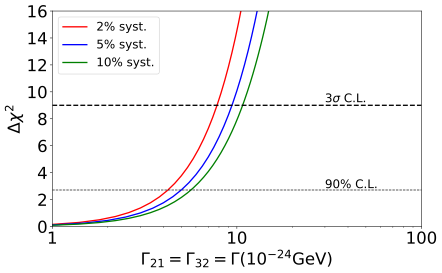}}
\caption{Sensitivity of the ESSnuSB experiment to quantum decoherence.}
\label{fig6}
\end{figure}

Fig.\ref{fig6} shows the sensitivity of the ESSnuSB facility to quantum decoherence using the main ESS beam and the FD \cite{ESSnuSB:2024yji}. This figure shows the capability of the ESSnuSB FD to constrain the decoherence parameter $\Gamma$. In this panel, different curves represent different values of the systematic error. In this analysis we have considered an open quantum system formalism where neutrino as a subsystem interacts with the environment giving rise to decoherence. While comparing our sensitivity with the other experiments, we find our sensitivity is comparable with DUNE. 

\subsection{Sensitivity to atmospheric neutrinos}

\begin{figure}[htb]
\centerline{
\includegraphics[width=15cm]{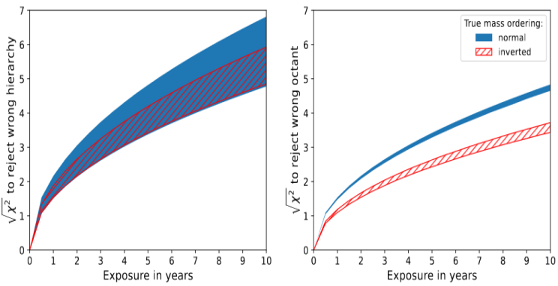}}
\caption{Sensitivity of the ESSnuSB experiment to atmospheric neutrinos.}
\label{fig7}
\end{figure}

Fig.\ref{fig7} shows the sensitivity of the ESSnuSB facility to atmospheric neutrinos that are produced in Earth’s atmosphere via interaction with the cosmic rays, using the the FD \cite{ESSnuSB:2024wet}. The left panel shows the capability of the ESSnuSB FD to reject the wrong neutrino mass ordering and the right panle shows the capability to reject the wrong octant of $\theta_{23}$ as a function of the exposure. In each panel the blue band is for normal mass ordering and the red band is for inverted mass ordering. In ESSnuSB FD we have excellent sensitivity as compared to other experiments due to the strong flux near the pole and the larger detector volume.

\subsection{Other sensitivities}

Apart from the results shown above, currently we are studying the sensitivity of the ESSnuSB FD to solar neutrinos and supernova neutrinos. Additionally study of sterile neutrinos involving the near detectors, LEMNB and LEnuSTORM is also in prgoress. The results are expected to be published soon.

\section{Summary and conclusion}
\label{sum}

 In this proceedings, we have shown the capability of the ESSnuSBplus setup to probe various physics scenarios in both standard 3 flavour and new physics cases. First we have outlined the experimental configuration of the ESSnuSBplus setup and then demonstrated the sensivity of the ESSnuSB far detector to CP violation and CP precision in the standard three flavour framework, sensitivity to sterile neutrinos, sensitivity to invisible neutrino decay, sensitivity to non-standard interaction mediated by scalar NSI, sensitivity to quantum decoherence and sensitivity to the atmospheric neutrinos. Estimation of the sensitivity of the far detector for solar neutrinos and supernova neutrinos and estimation of the sensitivity to sterile neutrinos using the near detectors END and LEMMOND and the neutrinos sources LEMNB and LEnuSRTORM are in progress. In conclusion we would like to stress that ESSnnuSB will be an extremely powerful neutrino facility in Europe with excellent sensitivities to a variety of physics cases.

\section*{Acknowledgements}

The work in part funded by (i) Ministry of Science and Education of Republic of Croatia grant No. KK.01.1.1.01.0001, (ii) Ministry of Science Education and Youth of the Republic of Croatia Grant No. PK.1.1.10.0002, (iii) SNSF and HRZZ under grant MAPS IZ11Z0$\_$230193 and (iv) European Union and European Union under the NextGenerationEU Programme. Views and opinions expressed are however those of the author(s) only and do not necessarily reflect those of the European Union. Neither the European Union nor the granting authority can be held responsible for them.

\end{document}